\documentclass[aip,jcp,reprint,superscriptaddress,floatfix]{revtex4-2}
\usepackage{graphicx}
\usepackage{dcolumn}
\usepackage{bm}
\usepackage[version=4]{mhchem}
\usepackage{siunitx}
\usepackage{xcolor}
\usepackage{placeins} 
\usepackage[super]{nth} 
\DeclareSIUnit{\um}{\micro\metre} 
\DeclareSIUnit{\wn}{\per\centi\meter} 
\DeclareSIUnit{\samples}{S}
\DeclareSIUnit{\molecule}{molecule}
\begin{document}
\preprint{APS/123-QED}

\title{Tracking water vapor homogeneous nucleation and droplet growth with spectroscopy and holography in a free expansion cloud chamber}

\author{Cole R. Sagan}
\affiliation{Department of Chemistry, Princeton University, Princeton, NJ 08544, USA}
\affiliation{Princeton Plasma Physics Laboratory, Princeton, NJ 08543, USA}

\author{Gwenore F. Pokrifka}
\affiliation{\mbox{High Meadows Environmental Institute, Princeton University, Princeton, NJ 08544, USA}}

\author{Samuel M. Koblensky}
\affiliation{\mbox{Department of Mechanical and Aerospace Engineering, Princeton University, Princeton, NJ 08540, USA}}

\author{Martin A. Erinin}
\altaffiliation[Current address: ]{Mechanical Engineering, University of Michigan, Ann Arbor, MI 48109, USA}
\affiliation{\mbox{Department of Mechanical and Aerospace Engineering, Princeton University, Princeton, NJ 08540, USA}}

\author{Ilian Ahmed}
\affiliation{\mbox{Department of Mechanical and Aerospace Engineering, Princeton University, Princeton, NJ 08540, USA}}
\affiliation{Département de Physique, École Normale Supérieure de Paris, Paris, France}

\author{Nadir Jeevanjee}
\affiliation{Geophysical Fluid Dynamics Laboratory, NOAA, Princeton, NJ 08540, USA}

\author{Luc Deike}
\email{ldeike@princeton.edu}
\affiliation{Princeton Plasma Physics Laboratory, Princeton, NJ 08543, USA}
\affiliation{\mbox{High Meadows Environmental Institute, Princeton University, Princeton, NJ 08544, USA}}
\affiliation{\mbox{Department of Mechanical and Aerospace Engineering, Princeton University, Princeton, NJ 08540, USA}}

\author{Marissa L. Weichman}
\email{weichman@princeton.edu}
\affiliation{Department of Chemistry, Princeton University, Princeton, NJ 08544, USA}
\affiliation{Princeton Plasma Physics Laboratory, Princeton, NJ 08543, USA}

\date{\today}
\begin{abstract}
We use a newly commissioned rapid expansion aerosol chamber (REACh) facility to study the homogeneous nucleation of water vapor to form liquid droplets.
We perform high-speed measurements to track the partitioning of water into vapor and droplets throughout the expansion process, including tunable diode laser absorption spectroscopy (TDLAS) to access the vapor concentration and in-line holography to track the size and concentration of nucleating droplets.
We retrieve the peak saturation ratio achieved in each expansion from the TDLAS measurements in combination with adjusted thermocouple temperature readout. 
We monitor the number of nucleated droplets and their subsequent growth as a function of saturation ratio, and observe the onset of homogeneous nucleation of water vapor occurring at a threshold saturation ratio near $S=5$, in agreement with prior literature and classical nucleation theory.
The trends we observe in average diameter and droplet concentration suggest that warm air pockets near the chamber walls inhomogeneously mix with cold air at the center of the chamber following expansion.
Active forced mixing with fans yields more spatially uniform temperature readings across the chamber, but also significantly broadens the droplet size distribution. 
Our results demonstrate the capability of TDLAS and holography techniques to track both water vapor and liquid water in the high saturation ratio environments necessary for the homogeneous nucleation of droplets.
Our findings also reveal that droplet nucleation and growth dynamics are highly sensitive to turbulence.
\end{abstract}

\maketitle

\section{Introduction} \label{sec:intro}

Cloud droplet formation occurs in two stages -- cluster nucleation and subsequent droplet growth -- which are governed by distinct physical mechanisms operating across vastly different length and time scales.\cite{lee2019new} 
Nucleation occurs at the microscopic level as individual vapor-phase molecules aggregate into thermodynamically-stable clusters. 
Beyond a critical size, the cluster can grow into a droplet, with growth dynamics transitioning from the free molecular regime ($r < \SI{10}{\nano\metre}$) to the diffusion controlled regime ($r > \SI{200}{\nano\metre}$).\cite{fladerer2003growth}
Throughout these processes, the evolution of a droplet's size is sensitive to environmental conditions, including turbulent mixing and temperature gradients that shape the local vapor field. 
These larger-scale effects can ultimately determine droplet size distributions in clouds.\cite{bodenschatz2010can, prabhakaran2020role, yeom2023cloud}
Understanding of droplet nucleation and growth under atmospheric conditions therefore requires one to account for the influence of the local vapor field, temperature gradients, and turbulence.
Here, we use the rapid expansion aerosol chamber (REACh) facility\cite{erinin2025droplet} equipped with high-speed optical diagnostics to track the homogeneous nucleation of water vapor and liquid droplet growth under controlled saturation ratio and turbulent mixing conditions. 

Homogeneous nucleation occurs when a supersaturated vapor spontaneously condenses into liquid droplets without any preexisting particles present to act as condensation nuclei.\cite{feder1966homogeneous,wyslouzil2016overview}
The homogeneous nucleation of water vapor in particular has been of fundamental interest for over a century. 
Pioneering experimental efforts by Wilson\cite{wilson1897condensation,wilson1899condensation} established expansion cloud chambers as a means to prepare supersaturated water vapor and investigate the conditions under which droplet nucleation occurs in the absence of condensation nuclei.
Next-generation expansion cloud chambers have been subsequently developed and applied to the homogeneous nucleation of water vapor throughout the 20$^\textrm{th}$ century\cite{powell1928condensation,volmer1934chamber,sander1943ubersattigung,allard1965new,wagner1981homogeneous,schmitt1981precision, strey1994problem,manka2007preliminary} 
and are still in wide use for a variety of applications.\cite{mohler2003experimental, wang2011design,tajiri2013novel,schnaiter2016cloud,erinin2025droplet} 
Homogeneous nucleation has also been studied in thermal diffusion cloud chambers\cite{brus2008homogeneous} and supersonic nozzles,\cite{hill1966condensation,paci2004spacially} though we focus here on expansion cloud chambers as they relate most closely to the instrumentation used in the present work.

Expansion cloud chambers are designed to mimic a rising air parcel in the atmosphere.
In a typical experiment, a parcel of gas experiences a sudden drop in pressure triggered via the motion of a piston or the opening of a valve into a secondary evacuated volume.
This pressure drop is accompanied by a drop in temperature and therefore a corresponding increase in the saturation ratio $S = e / e^{*}(T)$ where $e$ and $e^{*}(T)$ are, respectively, the partial pressure and temperature-dependent equilibrium vapor pressure of water vapor.
Homogeneous nucleation of water vapor is typically cited as occurring at saturation ratios exceeding $S_\mathrm{onset}= 4$ at \SI{293}{\kelvin},\cite{wilson1897condensation,volmer1934chamber,viisanen1993homogeneous,manka2007preliminary,wyslouzil2016overview} though this threshold depends strongly on a given experiment's diagnostic sensitivity to droplet formation -- both the lower size limit of droplet detection and the ability to detect individual droplets.
While $S$ rarely exceeds a few percent above 1 under typical atmospheric conditions, controlled laboratory experiments under more extreme conditions provide important comparison with theory, and a testbed for studying homogeneous ice nucleation.

Classical nucleation theory (CNT) provides a useful jumping-off point for the treatment of homogeneous nucleation.
Within CNT, individual vapor-phase monomers add to a small cluster until the cluster becomes thermodynamically stable and can evolve towards a macroscopic droplet.
The threshold between unstable and stable growth of a molecular cluster is set by the critical cluster radius, $r^*$, defined as:\cite{wyslouzil2016overview}
\begin{equation}
    r^* = \frac{2\gamma \nu_l}{k_{B} T \ln S}   \label{eq:rstar}
\end{equation}
where $\gamma$ is the surface tension of the liquid (\si{\newton\per\metre}), 
$\nu_l$ is the volume of one molecule (\si{\cubic\metre\per\molecule}), 
$k_B$ is the Boltzmann constant (\si{\joule\per\kelvin}),
and $T$ is temperature (\si{\kelvin}) taken to be equivalent for both the cluster and the surroundings, with heat generated by condensation assumed to equilibrate quickly.
The droplet nucleation rate $J$ (\si{\per\cubic\metre\per\second}) describes how often clusters will cross the $r^*$ critical size threshold at a given $T$ and $S$ and go on to form larger droplets:\cite{wyslouzil2016overview,becker1935kinetische}
\begin{equation}
    J = \sqrt{\frac{2\gamma}{\pi m}}\nu_l N^2 \;\exp\!\left[ -\frac{16\pi\gamma^3\nu_l^2}{3(k_B T)^3(\ln S)^2} \right]  \label{eq:J}
\end{equation}
where $m$ is the molecular mass (\si{\kilogram\per\molecule}) 
and $N$ is the vapor phase molecular number density (\si{\molecule\per\cubic\meter}).
Eq.~\ref{eq:J} assumes that the nucleation process is isothermal and that the surface tension of the spherical molecular cluster is comparable to that of the bulk liquid. 
The kinetic prefactor introduced by Becker and D\"{o}ring\cite{becker1935kinetische} assumes that the nucleation process is quasi-stationary.

Once a cluster surpasses the critical radius, sustained growth into a droplet can proceed.
The Hertz–Knudsen model describes the droplet growth rate in the free molecular regime as:\cite{hill1966condensation,pathak2013nonisothermal}
\begin{equation}
    \frac{dr}{dt} = \frac{\nu_l}{\sqrt{2\pi m k_{B}}}
    \left[ q_c\frac{e}{\sqrt{T}}-q_e\frac{e^*(\langle r \rangle,T_l)}{\sqrt{T_l}} \right]
    \label{eq:growth}
\end{equation}
where $r$ is the time-dependent droplet radius (\si{\meter}),
$T$ is the vapor-phase temperature (\si{\kelvin}),
$q_c$ and $q_e$ are dimensionless condensation and evaporation coefficients, 
$\langle r\rangle$ is the characteristic droplet radius in the ensemble (\si{\meter}),
and $T_l$ is the liquid-phase temperature (\si{\kelvin}), which is no longer assumed to be in equilibrium with the surrounding gas -- and is indeed expected to be elevated due to the heat of condensation.

Eqs.~\ref{eq:rstar}--\ref{eq:growth} make it clear that $S$ and $T$ are the most important environmental parameters governing homogeneous nucleation and droplet growth.  
Accurate measures of $S$ and $T$ are therefore critical during an expansion experiment.
Laboratory expansions are often assumed to be adiabatic, allowing one to express the maximum saturation ratio and minimum temperature reached during an experiment as simply:\cite{miller1983homogeneous,wilson1897condensation}
\begin{align}
    S_\mathrm{max,ad} &= \frac{e_{0}}{e^*(T_\mathrm{min,ad})} \cdot \frac{p_f}{p_0} \label{eq:S_ad} \\[2mm]
    T_\mathrm{min,ad} &= T_0 \left( \frac{p_{f}}{p_0} \right) \! ^{\frac{k - 1}{k}} \label{eq:T_ad}
\end{align}
where $e_0$, $p_0$, and $T_0$ are the initial partial pressure of water vapor, total chamber pressure, and  temperature before the expansion begins, 
$p_f$ is the final total chamber pressure after the expansion is complete, 
and $k = c_P / c_V$ is the ratio of specific heats of the background gas.

In practice, expansion experiments are never truly adiabatic. 
Various complications -- including expansion restriction by valves, spatial temperature gradients, evaporation of water from wet chamber walls, and thermalization with the walls -- introduce deviations from ideal behavior.\cite{schmitt1981precision}
These effects make it difficult to infer the actual peak saturation ratio $S_\textrm{max}$ reached in an expansion from $T_0$, $e_0$, and the $p_f/p_0$ expansion ratio alone.
In addition, the adiabatic approximation provides no description for how condensation of vapor during the expansion itself feeds back onto the evolution of $S$ and $T$, which in turn impact nucleation and subsequent droplet growth via Eqs.~\ref{eq:J} and ~\ref{eq:growth}.
To our knowledge, no prior experiments have simultaneously measured water vapor concentration and droplet evolution in real time during an expansion experiment near $S_\mathrm{onset}$ for the homogeneous nucleation of water vapor.
In addition, there have been no prior studies on the effect of turbulent mixing on water vapor homogeneous nucleation and droplet growth. 

Here, we use the REACh facility\cite{erinin2025droplet} to examine the homogeneous nucleation of water vapor, subsequent droplet growth, and the influence of turbulent mixing on these processes.
We deploy tunable diode laser absorption spectroscopy (TDLAS) to track the water vapor number density, inspired by prior demonstrations in chambers,\cite{ebert2005fibre,fahey2014aquavit,anderson2021effects} 
supersonic expansions,\cite{paci2004spacially,peng2019effective}
and aircraft measurements.\cite{buchholz2014rapid,buchholz2017hai} 
We use the TDLAS readout, in combination with adjusted thermocouple temperature measurements, to retrieve the maximum $S$ values reached during each expansion.
Concurrently, we harness in-line holography to track droplet concentrations and size distributions for droplets larger than \SI{8}{\micro\metre} in diameter.\cite{erinin2023comparison,erinin2025droplet}
We perform experiments in the absence of any seeding condensation nuclei to ensure that droplet formation occurs solely through homogeneous nucleation.


We find $S_\mathrm{onset} \simeq 5$ for homogeneous nucleation of water vapor at $T \simeq \SI{257}{\kelvin}$, in good agreement with CNT and prior experimental literature.\cite{miller1983homogeneous, viisanen1993homogeneous, manka2007preliminary}
We systematically consider expansions that are subcritical ($S_\mathrm{max}<4$), near-onset ($S_\mathrm{max} \sim 4-5$), and supercritical ($S_\mathrm{max}>5$) by varying the initial chamber saturation ratio and expansion ratio. 
We observe vapor depletion and droplet growth simultaneously, revealing tight coupling between droplet dynamics and the state of the surrounding vapor field.
We also assess how turbulent flow impacts the droplet size spectrum by monitoring water vapor fluctuations and droplet size distributions under different mixing conditions.
Forced mixing during the expansion appears to reduce large-scale ($\sim\SI{5}{\centi\metre}$) fluctuations in water vapor and temperature while also broadening the droplet size distribution, which we attribute to enhanced small-scale fluctuations.
With these findings, we establish TDLAS and holography as powerful tools to retrieve key statistics for droplet nucleation and growth. 
The present work will inform future studies tracking heterogeneous droplet nucleation and ice nucleation in the REACh facility.

\section{Experimental methods}  \label{sec:expt}

\begin{figure*}[tbp]
    \centering
    \includegraphics[width=\textwidth]{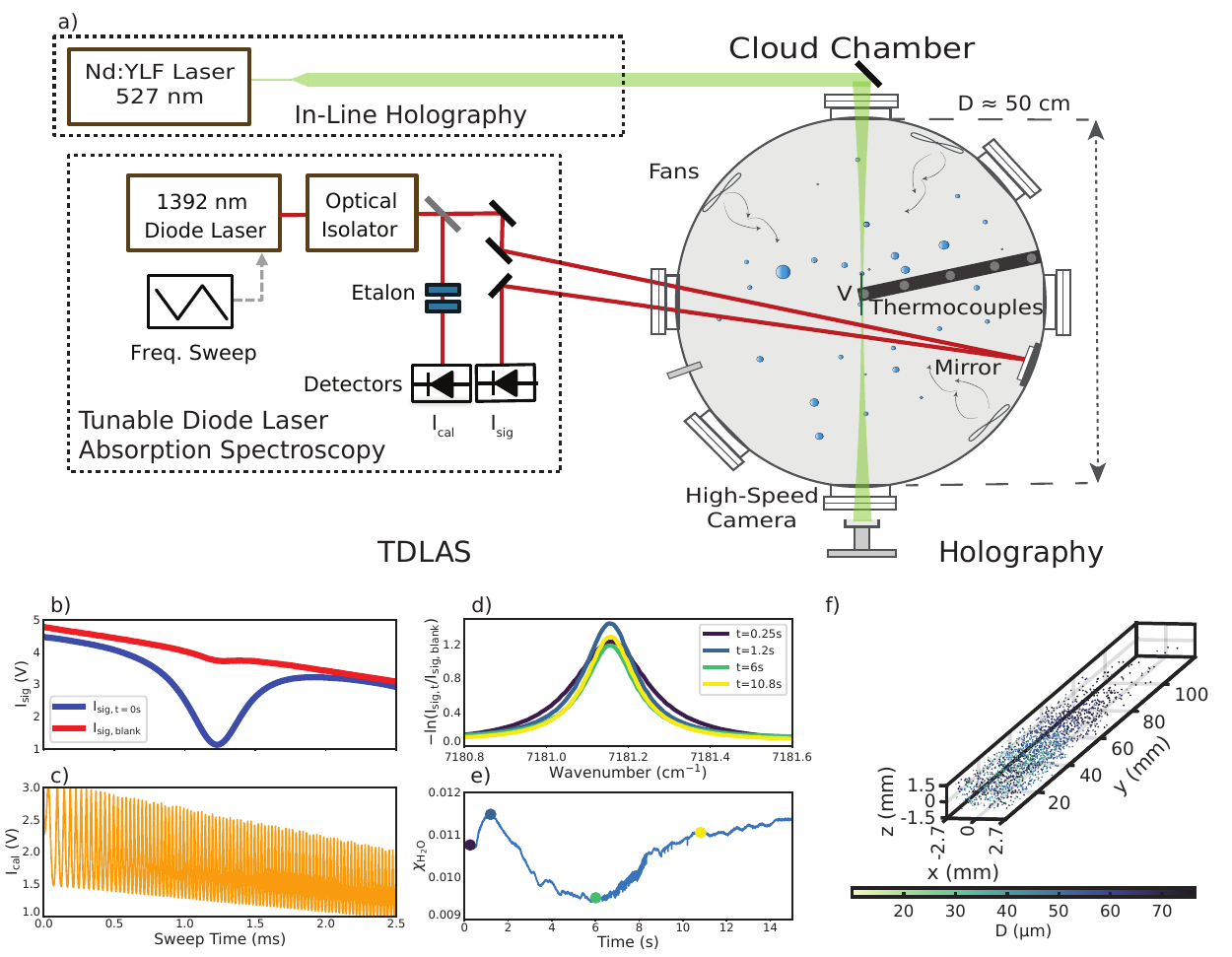}
    \caption{
    \textbf{(a)} 
    Schematic of the REACh facility. 
    Tunable diode laser absorption spectroscopy (TDLAS) is used to reconstruct the time-resolved water vapor mole fraction ($\chi_{\ce{H_2O}}$) inside the chamber, while time-synchronized in-line holography retrieves droplet concentrations and size distributions. 
    A high-speed thermocouple array measures the temperature inside the chamber during expansion and droplet growth. 
    Wall-mounted fans permit control over turbulence.
    \textbf{(b)}
    In the TDLAS measurements, the frequency of a diode laser operating near \SI{1392}{nm} is repeatedly swept through resonance with the $\nu_1+\nu_3, J=3\rightarrow2$ rovibrational transition of water vapor.  
    This laser light makes one round trip pass through the cloud chamber before its intensity is recorded with the signal detector. 
    $I_{\mathrm{sig},t=\SI{0}{s}}$ (blue trace) represents the signal beam intensity throughout the frequency sweep in the presence of water vapor at time $t=\SI{0}{s}$ before an expansion begins, with $S_0=0.5$ and $T_0=\SI{298}{K}$, as compared to the blank reference spectrum $I_\mathrm{sig,blank}$ (red trace) acquired with only dry \ce{N2} in the chamber.
    %
    \textbf{(c)}  
    To provide a relative frequency calibration for the TDLAS measurements, part of the \SI{1392}{nm} diode laser beam is picked off before the chamber, passed through a solid silicon etalon, and routed onto a second photodetector. 
    The intensity of light striking this photodetector, $I_\mathrm{cal}$, oscillates based on the known free spectral range of the etalon, allowing one to convert the sweep time axis to a frequency axis for plotting absorption spectra.
    \textbf{(d)} 
    Representative water vapor absorption spectra acquired at various times $t$ throughout an expansion.
    The area under the absorption peak reports directly on the water vapor concentration at each time step, while the absorption peak linewidth is sensitive to chamber pressure and, to a lesser degree, temperature.
    Data is shown for a representative expansion with $S_0=0.5$, $T_0=\SI{298}{K}$, and $\Delta p=\SI{390}{\hecto\pascal}$.
    \textbf{(e)} 
    The evolving mole fraction of water vapor $\chi_{\ce{H_2O}}$ can be reconstructed from the absorption spectra acquired throughout an expansion. 
    Absorption spectra are recorded with a repetition rate of \SI{400}{Hz}, corresponding to a sampling of $\chi_{\ce{H_2O}}$ every \SI{2.5}{\milli\second}. 
    Data is shown for an expansion with $S_0=0.5$, $T_0=\SI{298}{K}$, and $\Delta p=\SI{390}{\hecto\pascal}$, with color coded points corresponding to the absorption traces shown in panel (d).
    \textbf{(f)} 
    Reconstruction of spatial and size distributions of droplets formed within the holography sample volume aggregated over several images. 
    The axes show the spatial dimensions of the measurement volume in \si{\milli\metre}, and the color bar indicates the size of each detected droplet. 
    The sample volume for the largest droplets is indicated with the dark green line labeled \textit{V} in the center of the cloud chamber in panel (a). 
    }
    \label{fig:instrument}    
\end{figure*}

\subsection{Rapid expansion aerosol chamber (REACh) facility }

The REACh facility (Fig.~\ref{fig:instrument}a) is a free expansion cloud chamber consisting of two separate vacuum chambers -- the main cloud chamber where droplet nucleation takes place and a secondary expansion chamber connected via tubing and a solenoid valve (not pictured in Fig.~\ref{fig:instrument}a). 
The facility is described in detail in our previous report,\cite{erinin2025droplet} so we provide only a high-level overview here.

To prepare for an expansion experiment, we clean both chambers with several cycles of evacuation and purging with \ce{N2}.
We prepare the cloud chamber with initial saturation ratio $S_\mathrm{0}$ and total pressure $p_0$ by filling it with \ce{N2} carrier gas bubbled through ultrapure water (Milli-Q). 
The resulting \ce{H2O} and \ce{N2} mixture passes through a two-stage filter (La-Man 50 SCFM series, \SI{5}{\micro\metre} rating) to prevent small droplets from entering the chamber.
We confirm that the chamber is free of potential heterogeneous nucleation sites by sampling air from the chamber with a condensation particle counter (CPC, TSI 3007) which can detect particles down to $\SI{10}{\nano\metre}$.
This measurement yields a typical background particle concentration of $C_0=\SI{1}{\per\cubic\centi\metre}\pm{1}$ when the chamber is clean.

Once the cloud chamber is prepared with the desired initial conditions, the solenoid valve connecting it to the expansion chamber is opened to trigger an expansion and ultimately equalize the pressures across the two chambers. 
By varying $S_0$ and the pressure drop experienced in the cloud chamber, $\Delta p = p_0 - p_f$, we can access a wide range of $S_\mathrm{max}$ values during expansion. 
In all experiments reported here, the cloud chamber is initially prepared near atmospheric pressure, with $p_0 \approx\SI{1015}{\hecto\pascal}$. 
We therefore tune $\Delta p$ by varying via the initial level of vacuum in the secondary expansion chamber.
We measure $p_0$ and $p_f$ in the cloud chamber using a high-accuracy pressure transducer (Keller Preciseline High Accuracy Pressure Transmitter) with a \SI{5}{\hertz} acquisition rate.

The cloud chamber features a set of wall-mounted fans that can be turned on to induce turbulent flow.
The fans accelerate thermalization of the air with the chamber walls, which remain close to room temperature throughout an expansion.
Here, we report on experiments performed both with fans off (``unforced mixing'') and with fans on (``forced mixing'').
The flow in the chamber under these different mixing conditions has been characterized in our previous report.\cite{erinin2025droplet}
The unforced mixing experiments represent a closer analogue to expansion-style cloud chamber experiments in the prior literature. 

A home-made LABVIEW program interfaced with two data acquisition (DAQ) cards initiates data acquisition from the TDLAS setup, the holography camera system, and an array of thermocouple temperature sensors, and triggers each expansion. 
Data collection begins at $t=\SI{0}{\second}$ and the signal is sent to open the solenoid valve at $t=\SI{0.5}{\second}$, which then takes $\sim\SI{40}{\milli\second}$ to open before the expansion begins.
Expansions typically last for \SI{0.8}{s} before the chamber pressures equilibrate.
One DAQ card (National Instruments PCIe-6353) interfaced with a junction box (National Instruments BNC-2090A) samples the TDLAS photodetectors at a rate of \SI{250}{\kilo\samples\per\second}.
A second DAQ card (National Instruments PCIe-6320) interfaced with a second junction box (National Instruments SCB-68A) samples the thermocouple array at \SI{4}{\kilo\samples\per\second}, using the onboard cold junction compensation sensor for calibration. 
We discuss these measurements in detail in the following sections.

\subsection{Saturation ratio measurement with tunable diode laser absorption spectroscopy (TDLAS)} \label{sec:tdlas}

We use TDLAS to monitor the partial pressure of water vapor in the cloud chamber via the absorption of light by the $\nu_1 + \nu_3, J=3\rightarrow2$ rovibrational transition of water at \SI{7181.1557}{\wn} (\SI{1392.5335}{nm}), as shown in Fig.~\ref{fig:instrument}b–e.
The TDLAS system is based on a continuous-wave distributed feedback diode laser (Nanoplus TOP Wavelength \SI{1392}{nm}) fitted with an optical isolator to prevent optical feedback (Thorlabs IO-4-1390-VLP).
The TDLAS laser is centered at \SI{1392}{nm} (\SI{7184}{\wn}), with a $\sim$\SI{5}{\wn} tuning range, a sub-\SI{10}{MHz} instantaneous linewidth, and several milliwatts of output power.
We control the laser using a current driver (Thorlabs LDC202C) and a temperature controller (Thorlabs TTC001).
We modulate the laser wavelength by applying a \SI{200}{Hz} triangle wave from a waveform generator (BK Precision 4054b) to the analog modulation port of the current driver.
This modulation sweeps the laser frequency back and forth over a \SI{1.5}{\wn} window centered about the target rovibrational transition of water vapor, yielding TDLAS sampling every \SI{2.5}{\milli\second} during an experiment.

The TDLAS beam is split into two arms: a signal arm to quantify the amount of water vapor in the cloud chamber and a calibration arm aligned through an etalon to provide a relative frequency reference.
The calibration etalon traces are acquired synchronously alongside the water vapor absorption spectra from the signal arm throughout each expansion experiment. 
An enclosure purged with dry \ce{N2} surrounds the laser beamline outside  the chamber to minimize spurious signals from atmospheric water vapor.

In the signal arm, the \SI{1392}{nm} beam is aligned into the cloud chamber, back-reflected off a wall-mounted mirror, then collected and focused onto the signal photodetector (Thorlabs PDA50B2).
We first record a blank reference spectrum $I_\mathrm{sig,blank}$ when the chamber is filled with dry \ce{N2} (red trace in Fig.~\ref{fig:instrument}b).
We then acquire $I_{\mathrm{sig}}$ spectra throughout an expansion; 
a representative trace is shown in blue in Fig.~\ref{fig:instrument}b, acquired at time $t=\SI{0}{\second}$ before an expansion begins, with $S_0 = 0.5$ and $T_0 = \SI{298}{\kelvin}$. 
The signal attenuation due to resonant absorption of light by water vapor in the chamber is evident.

In the parallel calibration arm, \SI{1392}{nm} light passes through a solid silicon Fabry-P\'{e}rot etalon with a free spectral range (FSR) of $\sim$\SI{0.5}{GHz} (Light Machinery OP-2638-83110) and strikes the calibration photodetector (Thorlabs PDA30B2).
Figure~\ref{fig:instrument}c shows the oscillating laser intensity ($I_\mathrm{cal}$) transmitted through the etalon and incident on this detector as the laser frequency is swept.
We calibrate the FSR by reference to three rovibrational transitions of water vapor whose frequencies are available in the HITRAN database.\cite{HITRAN}
The positions of the maxima and minima of each $I_\mathrm{cal}$ trace are extracted and mapped to frequencies spaced by half the etalon FSR using a quadratic fitting function.
This calibration procedure allows us to convert the laser sweep time axis to a frequency axis for the water vapor absorption spectra, while correcting for any non-linearity in the laser sweep, yielding $I_\mathrm{sig}(\nu)$ and $I_\mathrm{sig,blank}(\nu)$.

At each time point, we calculate the absorbance, $A(\nu)$, by assuming that $I_{\mathrm{sig}}(\nu)$ is attenuated with respect to $I_\mathrm{sig,blank}(\nu)$ according to the Beer-Lambert law:
\begin{equation}
    A(\nu) = -\ln\left[ \frac{I_{\mathrm{sig}}(\nu)}{I_\mathrm{sig,blank}(\nu)}\right] =  \sigma(\nu) N L 
    \label{eq:abs}
\end{equation}
%
where $\sigma(\nu)$ is the molecular absorption cross section (\si{\metre\squared\per\molecule}), 
$N$ is the number density of absorbing molecules in the chamber (\si{\molecule\per\cubic\metre}), 
and $L=\SI{0.97}{m}$ is the pathlength of light through the chamber.
Absorption spectra are shown in Fig.~\ref{fig:instrument}d for various time points in a representative expansion with $S_0=0.5$ and $\Delta p=\SI{390}{\hecto\pascal}$.

The molecular absorption cross section is given by:
\begin{align}
    \sigma(\nu) = S_{ij}(T) \cdot f(\nu; T,p) 
    \label{eq:sigma}
\end{align}
where $S_{ij}(T)$ is the temperature-dependent spectral line intensity of the relevant rovibrational transition (\si{\metre\per\molecule}), available for water vapor from HITRAN,\cite{HITRAN,lisak2009spectroscopic} and $f(\nu; T,p)$ is the normalized temperature- and pressure-dependent spectral lineshape function (\si{\metre}). 
Here, $f(\nu; T,p)$ is taken to be a Voigt profile arising from the convolution of Lorentzian pressure broadening and Gaussian Doppler broadening, though pressure broadening dominates under our conditions. 
We fit each absorption spectrum using the Voigt peak profile found in the Python lmfit package,\cite{lmfit} floating the area under the absorption peak and the Lorenzian half-width at half-maximum (hwhm) linewidth to account for changing water vapor content and pressure in the chamber throughout the expansion.
During each fit, we fix the Gaussian Doppler linewidth using the average (unadjusted) temperature measured by the thermocouple array in the chamber at time $t$; see Section \ref{sec:temp} below for further details on temperature measurements.

Combining Eqs.~\ref{eq:abs} and \ref{eq:sigma} and integrating over the spectral axis, one can use $a$, the area under the experimental absorption curve (\si{\per\meter}) at each time point, to find the molecular number density:
\begin{equation}
    N = \frac{a}{S_{ij}(T) \, L} \label{eq:N}
\end{equation}
%
%
from which we can calculate the partial pressure of water vapor, $e$, using the ideal gas law:
\begin{align}
    e = N k_B T\label{eq:eTDLAS}
\end{align}
where $T$ is the average unadjusted temperature read out via the thermocouple array at each sampled time point (see Section \ref{sec:temp} below).

In addition, the hwhm pressure-broadened Lorentzian linewidth of the water vapor lines measured with TDLAS, $\gamma_L$ (cm$^{-1}$), is related to the total chamber pressure, $p$, and partial pressure of water vapor, $e$, by:
\begin{equation}
    \gamma_L(p,T)=\left(\frac{T_\mathrm{ref}}{T}\right)^{n_\mathrm{air}}
    \Big[ \gamma_\mathrm{air}(p-e)+\gamma_\mathrm{self} \, e \Big] 
    \label{eq:pressure_broadening}
\end{equation}
where $n_\textrm{air}$ is a unitless coefficient describing the temperature dependence of the pressure broadening, $\gamma_\textrm{air}$ is the pressure broadening coefficient for collisions with air (cm$^{-1}$/atm), and $\gamma_\textrm{self}$ is the pressure broadening coefficient for collisions with other water molecules (cm$^{-1}$/atm). 
These quantities are available for the water vapor transitions of interest from the HITRAN database\cite{HITRAN} and defined at reference temperature $T_\textrm{ref}=\SI{296}{K}$.
TDLAS readout of the water vapor transition linewidths therefore permits a rapid, contactless measurement of $p$ with a repetition rate synced with the absorption measurements of $N$.
The pressures obtained \textit{via} TDLAS are consistently $\sim3\%$ higher than those obtained using high accuracy pressure sensors, though expansion ratios are in near exact agreement. 
As such, we use the dedicated sensors for high-accuracy static readout of chamber pressure, but use the TDLAS readout requiring real-time pressure readings throughout an expansion. 

With $e$ and $p$ in hand via Eqs.~\ref{eq:eTDLAS} and \ref{eq:pressure_broadening}, we can calculate 
mole fraction of water vapor, $\chi_{\ce{H2O}}$ as:
\begin{align}
    \chi_{\ce{H2O}} &= e / p     \label{eq:mol_frac}
\end{align}
%
This process ultimately yields the path averaged evolution of $\chi_{\ce{H2O}}$ due to condensation and evaporation sampled at \SI{400}{Hz} throughout each expansion. 
Figure~\ref{fig:instrument}e shows the evolution in $\chi_{\ce{H2O}}$ over the course of a single representative expansion with $S_0=0.5$ and $\Delta p=\SI{390}{\hecto\pascal}$.
We harness TDLAS as a powerful \textit{in situ} measurement of water vapor content and chamber pressure during homogeneous nucleation expansion experiments in the following results and discussion.

\subsection{Time-resolved thermocouple temperature measurements \& temperature correction} \label{sec:temp}
We use an array of highly-responsive type-T thermocouple sensors (Omega COCO-002) to track temperature in the cloud chamber during an expansion with some degree of spatial resolution at a sample rate of \SI{4}{kHz}.
One freely suspended thermocouple sensor is mounted roughly in the center of the chamber. 
Another five sensors are evenly spaced \SI{6}{cm} apart in an array mounted on a plastic holder that lies \SI{5}{cm} from the top of the chamber (see Fig.~\ref{fig:instrument}a). 
One end of this sensor array lies near the center of the chamber while the outermost sensor lies within \SI{2}{cm} of the chamber wall. 
The freely suspended thermocouple typically reads the lowest temperatures, while the thermocouple closest to the wall of the chamber has significantly warmer readings, typically $\sim\SI{20}{K}$ higher than the coldest thermocouple.
As detailed in Section \ref{sec:tdlas} above, calculations involving TDLAS measurements that require fast sampling in time make use of the raw, unadjusted temperature averaged across the thermocouple array, including the suspended thermocouple, but excluding the warmest thermocouple mounted closest to the wall.
However, for any calculations that deal with the minimum value of $T$ or maximum value of $S$ reached during an expansion (\textit{vide infra)}, we use the temperature reading from the single coldest thermocouple (usually the freely suspended sensor), which we adjust to account for latent heat of condensation on the thermocouple itself.

It is important to use sufficiently responsive thermocouples to measure the transient minimum temperatures reached during an expansion.
In order for a thermocouple to have a quick response rate, it must be highly thermally conductive, i.e.\ uncoated and with very small thermal mass. 
However, highly responsive thermocouples also add certain experimental complications in high $S$ environments like the REACh cloud chamber.
In particular, the latent heat release from the condensation of a small amount of water vapor on a thermocouple itself can significantly heat up the sensor, leading to artifacts in temperature readout.
In order to more accurately report temperature during the expansion, we adjust our temperature measurements to account for water condensation on the sensors.
This temperature correction procedure is described in detail in Section SI of the Supplementary Material (SM).
These corrections allow us to accurately retrieve the minimum temperature reached during expansion, which we can then use to calculate the peak saturation ratio, $S_\mathrm{max}$, according to:
\begin{equation}
    S_\mathrm{max} = S_\mathrm{max,exp} \cdot \frac{e^*(T_\mathrm{min,exp})}{e^*(T_\mathrm{min,adj})} %
    \label{eq:S_adj}
\end{equation}
where $S_\mathrm{max,exp}$ is the maximum saturation ratio calculated using $e$ from experimental TDLAS measurements and the unadjusted thermocouple reading, and $e^*(T_\mathrm{min,exp}$) and $e^*(T_\mathrm{min,adj}$) are the equilibrium vapor pressures of water at the minimum unadjusted and adjusted temperatures reached in the expansion, respectively.
In the following results, $S_\mathrm{max}$ represents a key parameter that determines the homogeneous water droplet nucleation rate.

\subsection{Droplet growth measurements with in-line holography} \label{sec:holography}
In parallel to the water vapor measurements described above, we use in-line holography to track water droplet concentrations and size distributions in time.
Holography can also be used to track the spatial distributions of droplets, as illustrated in Fig.~\ref{fig:instrument}f, though we do not make use of this capability here.
Details of the holography system are provided by \citet{erinin2023comparison}
In brief, the system utilizes collimated light from an Nd:YFL laser (CrystalLaser, QL527-200-L), the diffraction of which is recorded by a 4K camera  (Phantom, VEO4k-990-L) with a long-range microscope (Infinity Photo-Optical Co., K2 DistaMax). 
The system features a measurement repetition rate of \SI{600}{Hz} and
allows for the detection of droplets down to \SI{8}{\micro\metre}. 

Each captured hologram is reconstructed in depth to locate and measure droplets within a 3D reconstruction volume.
This measurement volume depends on the camera sensor size and lens magnification, and it varies linearly with particle size.\cite{nakai2025particleholographyjl} 
The volume is calibrated for each setup using resolution targets.
One exemplary set of calibrated values features a cross-sectional area of $\SI{0.54}{cm} \times \SI{0.31}{cm}$. 
With this field of view, a $\SI{10}{\micro\metre}$ droplet and a $\SI{50}{\micro\metre}$ droplet have respective reconstruction depths of \SI{3.31}{cm} and \SI{16.55}{cm} along the chamber diameter, granting corresponding total measurement volumes of $\SI{0.55}{cm^3}$ and $\SI{2.77}{cm^3}$. 
All experiments described herein have comparable measurement volumes.

Holography images are acquired for \SI{9}{\second}, beginning $\SI{0.5}{\second}$ before the expansion is triggered. 
The holographic processing is performed in two steps, as described by \citet{erinin2023comparison}. 
In the first step, droplet locations and approximate diameters are found for each captured hologram. 
In the second step, each located drop is individually re-processed to determine a precise diameter and to remove false detections. 
Due to computational expense, the datasets that underpin Figs.~\ref{fig:J_v_Smax} and \ref{fig:J_drops} are processed through the first step, and the second refinement step is performed only for a \SI{0.33}{s} time window centered about $t = \SI{3}{\second}$, where the maximum concentration of droplets typically occurs.
For the eight expansion experiments whose full time-series are presented here in Figs.~\ref{fig:ABC}--\ref{fig:drop_percent}, holography data is processed through both steps for over the entire \SI{9}{\second} time interval.

Unless otherwise noted, we process the droplet statistics data as follows.
For each holography image, the size distribution of droplets is measured, binned, and normalized by the bin size and measurement volume at the mean diameter of that bin. 
This yields time-dependent size distributions reported as $N_d(D,t)$ in units of \si{\per\micro\metre\per\cubic\centi\metre}.
$N_d(D,t)$ is averaged in time in intervals of 9 holography images, granting distribution measurements every \SI{1.5}{\milli\second} during an expansion.
We calculate the droplet concentration, $C_\textrm{drop}$ (\si{\per\cubic\centi\metre}), by integrating the size distribution over all diameters:
\begin{equation}
    C_\textrm{drop} = \int_0^\infty N_d(D,t)~\mathrm{d}D %
\end{equation}
The maximum droplet concentrations reported correspond to the presence of \SIrange{2}{70}{} droplets in a single frame.
We report the average diameter, $\overline{D}$ (\si{\micro\metre}), by taking the first moment of the size distribution over all diameters, normalized by the droplet concentration:
\begin{equation}
    \overline{D} = \frac{\int_0^\infty N_d(D,t)D~\mathrm{d}D}{C_\textrm{drop}} 
\end{equation}

At sufficiently high droplet concentrations, tracking and retrieval of droplets by holography becomes intractable due to optical interference from droplets both above and below the \SI{8}{\micro\metre} holography detection size threshold.
In experiments studying the homogeneous nucleation of water vapor in particular, we expect the formation of many small droplets which cannot be reconstructed accurately with holography but nonetheless cause optical interference.
To mitigate these effects, we neglect all droplets with diameters below $\SI{8}{\micro\metre}$ since they are difficult to distinguish from noise. 
Visually, the optical interference starts to become significant only for the highest concentrations we measure, $C_\textrm{drop}>$ \SI{50}{\per\cubic\centi\metre}. 
These conditions are expected to suffer from higher measurement uncertainties. 
Conditions where interference significantly impacts the droplet detections are therefore removed from our analysis. 
Videos of holography data recorded in expansion experiments with significant interference from high $C_\textrm{drop}>$ at high $S_\mathrm{max}$ as compared to expansion that reach reasonable $C_\textrm{drop}>$ closer to $S_\mathrm{onset}$ are available in the SM (Video~S1 and Video~S2, for the interference and clean cases, respectively).

\section{Results and Discussion}

We now present the results from expansion experiments spanning various conditions. 
We first consider the concentration and size distributions of homogeneously nucleated water droplets as a function of $S_\mathrm{max}$ and compare the findings with CNT.
We then consider droplet growth dynamics using time-resolved data from both TDLAS and holography.
Finally, we discuss the impact of turbulent flow on homogeneous nucleation and droplet growth by using fans to force mixing inside the chamber.

\subsection{Homogeneous nucleation of water vapor} \label{sec:nuc}

\begin{figure}
\includegraphics[width=\columnwidth]{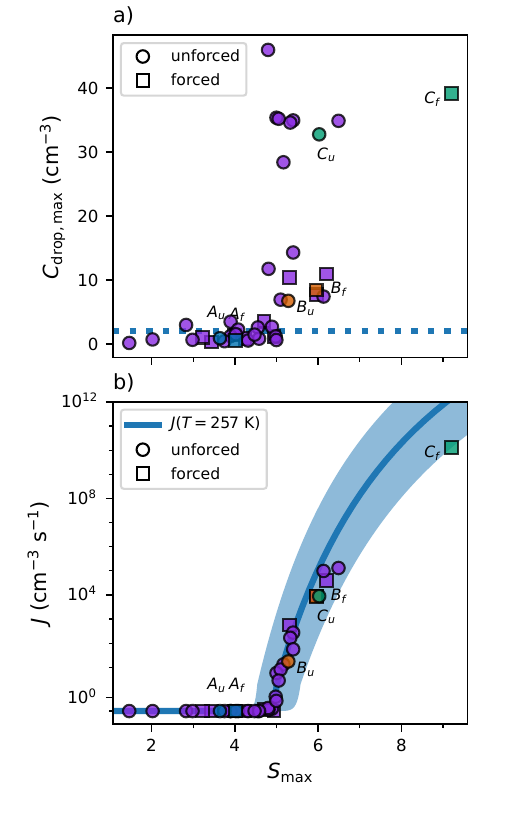}
   \caption
    {
    \textbf{(a)} 
    Maximum concentration of droplets measured during an expansion experiment vs.\ peak saturation ratio reached in that expansion.
    $C_\mathrm{drop,max}$ is measured with holography as described in Section \ref{sec:holography} while $S_\textrm{max}$ is calculated from TDLAS and thermocouple data using Eq.~\ref{eq:S_adj}.
    Forced mixing experiments (squares) have fans on inside the chamber, while unforced mixing experiments (circles) have fans off.
    Points labeled A, B, and C correspond to experiments whose time series are detailed in Figs.~\ref{fig:ABC} and \ref{fig:ABC_fans}.
    The horizontal dashed line marks a threshold of $C_\mathrm{drop, max} = {\SI{2}{cm^{-3}}}$ above which homogeneous nucleation is clearly occurring.
    \textbf{(b)} 
    Homogeneous nucleation rate $J$ predicted for each experiment using Eq.~\ref{eq:J} and the experimental values for $S_\mathrm{max}$ and $T_\mathrm{min,adj}$.
    The solid blue line plots $J$ calculated as a function of $S_\mathrm{max}$ at $T=\SI{257}{\kelvin}$, the average $T_\mathrm{min,adj}$ across all experiments, with the blue envelope showing the spread in $J$ from the minimum and maximum $T_\mathrm{min,adj}$ values.
    }
    \label{fig:J_v_Smax}
\end{figure}

We begin by considering holography data from a series of expansion experiments performed with different $S_\mathrm{0}$ and $\Delta p$.
The conditions for all expansions performed are summarized in Table~S1 of the SM.
%
Figure~\ref{fig:J_v_Smax}a plots the maximum concentration of droplets reached any point in time during an expansion experiment, $C_\mathrm{drop,max}$, versus $S_\mathrm{max}$ for that expansion calculated with Eq.~\ref{eq:S_adj}. 
Experiments performed with unforced (fans off) mixing conditions are notated with circles, 
while forced (fans on) mixing experiments are plotted with squares.
We target conditions where $S_\mathrm{max}$ lies well below the homogeneous nucleation threshold (blue points marked A), near the homogeneous nucleation threshold (orange points marked B), and well above it (green points marked C). The subscripts for the A, B, and C labels correspond to whether the experiment features forced (f) or unforced (u) mixing.
Time-resolved data from these same sets of experiments is detailed further in Section \ref{sec:growth}.

$C_\mathrm{drop,max}$ exceeds $\SI{2}{cm^{-3}}$ (dotted blue line in Fig.~\ref{fig:J_v_Smax}a) for $S_\mathrm{max}>5$, indicating that the onset of homogeneous nucleation lies near this saturation ratio, e.g.\ $S_\mathrm{onset}\simeq5$.
We take $\SI{2}{cm^{-3}}$ as a conservative threshold that lies above the background particle concentration, and corresponds to a statistically significant number of droplets in the measurement volume.
$C_\mathrm{drop,max}$ continues to increase dramatically for $S_\mathrm{max} > S_\mathrm{onset}$, peaking at \SI{46}{cm^{-3}}.
The inability of holography to detect particles smaller than $\SI{8}{\micro\metre}$ in diameter likely leads to under-reporting of $C_\mathrm{drop,max}$ at higher $S_\mathrm{max}$.
The spread in $C_\mathrm{drop,max}$ for a given $S_\mathrm{max}$ arises from the fact that two expansions with distinct $\Delta p$ and $S_0$ may happen to reach the same $S_\mathrm{max}$ with different values of $e$ and $T$.
Inspection of Eq.~\ref{eq:J} makes it clear that different combinations of $S$ and $T$ might yield vastly different homogeneous nucleation rates $J$, and therefore different $C_\textrm{drop}$.

Figure \ref{fig:J_v_Smax}b plots the homogeneous nucleation rate $J$ calculated with Eq.~\ref{eq:J} for each data point in Fig.~\ref{fig:J_v_Smax}a, using the experimental $S_\mathrm{max}$ and $T_\mathrm{min,adj}$ for each expansion.
The data now collapses nicely to a line as compared to Fig.~\ref{fig:J_v_Smax}a.
We calculate three different $J$ isotherms to account for the range of temperatures reached in different expansions. 
Across all experiments where homogeneous nucleation occurs and $C_\mathrm{drop,max}$ exceeds the $\SI{2}{cm^{-3}}$ threshold, the average $T_\mathrm{min,adj}$ reached is $\SI{257}{\kelvin}$.
The solid blue line in Fig.~\ref{fig:J_v_Smax}b plots $J$ calculated as a function of $S_\mathrm{max}$, with $T$ fixed at $\SI{257}{\kelvin}$.
Many experiments that feature $5 < S_\mathrm{max} <6$ reach a similar minimum temperature and fall on this isotherm.
We repeat this calculation of $J$ using the minimum and maximum $T_\mathrm{min,adj}$ values reached across all experiments to yield the bounds of the shaded blue envelope in Fig.~\ref{fig:J_v_Smax}b.
All experimental results fall well within these bounds.
In addition, the calculated $J$ exceeds \SI{1}{\per\cubic\centi\metre\per\second} for $S_\mathrm{max} > 5$, in close agreement with the results of Fig.~\ref{fig:J_v_Smax}a.
We can now use these calculated $J$ values to understand how the concentration and typical size of nucleated droplets respond to changing nucleation rates.

\begin{figure}[tbp]
    \centering
    \includegraphics[width=\columnwidth]{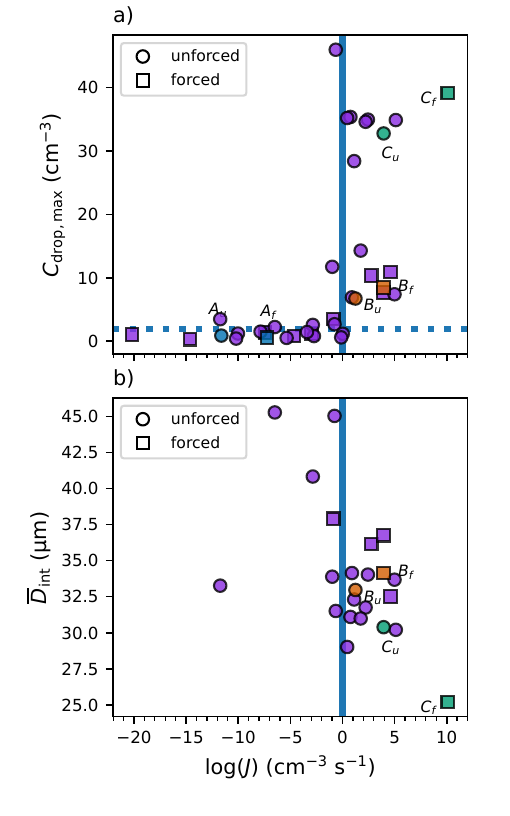}
    \caption{
    \textbf{(a)} Maximum droplet concentration, $C_\mathrm{drop,max}$ and \textbf{(b)} average droplet diameter $\overline{D}_\textrm{int}$ measured with holography as a function of the homogeneous nucleation rate $J$ calculated for each expansion using Eq.~\ref{eq:J} with experimental $S_\mathrm{max}$ and $T_\mathrm{min,adj}$ values. 
    $\overline{D}_\textrm{int}$ is calculated over the \SI{0.33}{\second} hologram reconstruction time interval centered about $t=\SI{3}{\second}$, when $C_\mathrm{drop}$ typically reaches its peak. 
    The horizontal dashed line in panel (a) marks the $\SI{2}{cm^{-3}}$ threshold for detection of homogeneously nucleated droplets.
    The vertical blue line in both panels marks $J=\SI{1}{\per\cubic\centi\metre\per\second}$, above which we expect to observe detectable concentrations of nucleated droplets.
    Points labeled A, B, and C correspond to the same experiments marked in Fig.~\ref{fig:J_v_Smax} whose time series are detailed in Figs.~\ref{fig:ABC} and \ref{fig:ABC_fans}.
    }
    \label{fig:J_drops}
\end{figure}

Figure \ref{fig:J_drops}a plots $C_\mathrm{drop,max}$ as a function of calculated $J$ for the same set of experiments from Fig.~\ref{fig:J_v_Smax}.
Nearly all experiments with $C_\mathrm{drop,max}>${\SI{2}{\per\cubic\centi\metre}} (threshold marked with a horizontal blue dotted line) feature $J>\SI{1}{\per\cubic\centi\metre\per\second}$ (threshold marked with a vertical blue line).
We stress that we find only qualitative agreement between the rate of nucleation and the count of nucleated droplets: while the full expansion lasts \SI{0.8}{\s}, we cannot be certain of the exact time window during which the homogeneous nucleation rate is appreciable.
For $J>\SI{1}{\per\cubic\centi\metre\per\second}$, $C_\mathrm{drop,max}$ increases with $J$ to a point, then levels off.


We also consider the average size of droplets formed in each expansion. 
Figure \ref{fig:J_drops}b plots the average droplet diameter $\overline{D}_\textrm{int}$ measured with holography versus $J$.
$\overline{D}_\textrm{int}$ is calculated over the time interval when the number of homogeneously nucleated droplets reaches a maximum, in a \SI{0.33}{\second} window centered about $t=\SI{3}{\second}$. 
We calculate $\overline{D}_\textrm{int}$ only for experiments where $C_\textrm{drop,max}>\SI{2}{cm^{-3}}$ in order to exclude background particles and very low droplet-count experiments from the analysis.
The general trend observed is that $\overline{D}_\textrm{int}$ decreases as $J$ increases.
We do observe some spread in $\overline{D}_\textrm{int}$ for a given $J$, as $S_0$ can vary widely across experiments, changing the amount of water vapor available for condensation.

We can explain the results in both panels of Fig.~\ref{fig:J_drops} by noting that as the concentration of homogeneously nucleated droplets increases with $J$, the competition for water vapor increases and limits the size to which a given droplet can grow.
Hence, $\overline{D}_\textrm{int}$ decreases with $J$.
In addition, at high $J$ we likely form a large number of very small droplets that fall below the holography detection limit, leading the recorded $C_\mathrm{drop,max}$ to appear to level off.
These too-small droplets can also cause optical interference in the holography measurement, limiting our ability to retrieve accurate droplet statistics at high $J$ and likely leading to additional scatter in $C_\mathrm{drop,max}$.
%

\subsection{Droplet growth under unforced mixing conditions} \label{sec:growth}

\begin{figure}
    \includegraphics[width=\columnwidth]{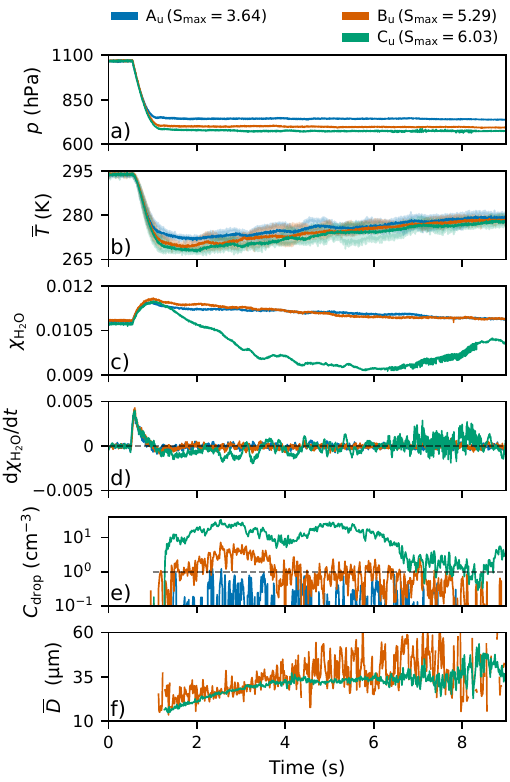}
    \caption{
    Evolution of water vapor and droplet statistics over the course of three expansions performed under unforced mixing conditions (fans off) with $S_0=0.50$, corresponding to the color-coded circle markers in Figs.~\ref{fig:J_v_Smax} and \ref{fig:J_drops}.
    Experiment A$_\mathrm{u}$ (blue traces) features $\Delta p=\SI{329}{\hecto\pascal}$, and reaches $S_\mathrm{max}=3.64$, below $S_\mathrm{onset}$.   
    Experiment B$_\mathrm{u}$ (orange traces) features $\Delta p=\SI{374}{\hecto\pascal}$, and reaches $S_\mathrm{max}=5.29$, near $S_\mathrm{onset}$.
    Experiment C$_\mathrm{u}$ (green traces) features $\Delta p=\SI{390}{\hecto\pascal}$, and reaches $S_\mathrm{max}=6.03$, above  $S_\mathrm{onset}$.
    \textbf{(a)} Chamber pressure retrieved from linewidth fits of TDLAS water vapor absorption lines. 
    \textbf{(b)} Unadjusted average temperature reading across the thermocouple array; the shaded envelopes show the coldest and warmest thermocouple readings at each time point.
    \textbf{(c)} Mole fraction of water vapor measured with TDLAS. 
    \textbf{(d)} $\mathrm{d}\chi_{\ce{H2O}}/\mathrm{d}t$ calculated numerically after smoothing the traces in panel (c). 
    \textbf{(e)} Droplet concentration tracked with holography.
    The dashed black line indicates the droplet concentration threshold of $\SI{2}{cm^{-3}}$ that distinguishes homogeneously nucleated droplets from activation of background particles.
    \textbf{(f)} The ensemble average diameter of all droplets counted across  9 consecutive frames tracked with holography.
    We exclude A$_\mathrm{u}$ from this panel, as the low droplet counts make averaging impractical.
    }
    \label{fig:ABC}
\end{figure}

We now make use of the high time resolution of TDLAS and holography to understand the evolution of the water vapor field and droplet growth for different $S_\textrm{max}$ and turbulence conditions.
Figure~\ref{fig:ABC} reports droplet formation statistics under unforced (fans off) mixing conditions for the expansions labeled A$_\mathrm{u}$, B$_\mathrm{u}$, and C$_\mathrm{u}$ in Figs.~\ref{fig:J_v_Smax} and \ref{fig:J_drops}.
%
%
All three of these experiments feature initial $S_\mathrm{0}=0.5$, with $S_\mathrm{max}$ tuned by changing $\Delta p$.

Figure~\ref{fig:ABC}a plots the cloud chamber pressure as a function of time to illustrate the timing of the expansion and the final pressure reached in each experiment.
Figure~\ref{fig:ABC}b plots the unadjusted average thermocouple array temperature (solid lines) as a function of time, with the envelopes bounded by the warmest and coldest temperatures measured in the array at each time.
Figure~\ref{fig:ABC}c tracks the water vapor mole fraction, $\chi_{\ce{H2O}}$, throughout the expansion recovered using the TDLAS measurements as described in Section \ref{sec:tdlas}.
We also plot the first time derivative of $\chi_{\ce{H2O}}$ in Figure~\ref{fig:ABC}d to better emphasize how water vapor is appearing ($\mathrm{d}\chi_{\ce{H2O}}/\mathrm{d}t >0$) and disappearing ($\mathrm{d}\chi_{\ce{H2O}}/\mathrm{d}t <0$) throughout the experiment.
Figures~\ref{fig:ABC}e,f show the evolution of droplet concentration, $C_\mathrm{drop}$, and mean droplet diameter, $\overline{D}$, retrieved with holography.

Experiment A$_\mathrm{u}$ features $S_\mathrm{max}<5$ and therefore corresponds to a regime in which background particles in the chamber are activated to form droplets, but no homogeneous nucleation of water vapor is expected.
Accordingly, holography measures $C_\mathrm{drop}$ well below the \SI{2}{cm^{-3}} threshold marked with a black dashed line in Fig.~\ref{fig:ABC}e.

Experiments B$_\mathrm{u}$ and C$_\mathrm{u}$ feature larger pressure drops to reach lower temperatures and therefore higher $S_\mathrm{max}$. 
B$_\mathrm{u}$ and C$_\mathrm{u}$ reach droplet concentrations of \SI{6.7}{\per\cubic\centi\metre} and \SI{33}{\per\cubic\centi\metre}, respectively, appreciably higher than the background particle concentration, and therefore a clear indication that homogeneous nucleation is occurring.
The average droplet diameter $\overline{D}$ is also significantly higher in experiment B$_\mathrm{u}$ than it is in C$_\mathrm{u}$ (see Fig.~\ref{fig:ABC}f).
This is expected, as the average droplet diameter should decrease when more droplets are formed given fixed available water vapor, as also discussed above in Section \ref{sec:nuc}.

We now consider the trends in water vapor mole fraction in Fig.~\ref{fig:ABC}c,d, whose evolution is closely linked with droplet formation and growth.
Experiments A$_\mathrm{u}$, B$_\mathrm{u}$, and C$_\mathrm{u}$ all show a clear initial increase in $\chi_{\ce{H2O}}$ at the start of the expansion, also evident as a sharp positive jump in $\mathrm{d}\chi_{\ce{H2O}}/\mathrm{d}t$.
We ascribe this increase in $\chi_{\ce{H2O}}$ to the evaporation of a film of liquid water that adsorbs on the walls during the pre-expansion humidification of the chamber.
The rapid pressure drop at the start of the experiment pulls this water off the walls and into the vapor phase.
Taking the jump in $\chi_{\ce{H2O}}$ as approximately +0.005, and accounting for the volume and surface area of the cloud chamber, we estimate that the evaporating film is $\sim\SI{50}{nm}$ thick. 
This estimate is in reasonable agreement with prior work that has observed similar phenomena.\cite{dobrozemsky1995reduction, dobrozemsky2007residence, sefa2013study}

\begin{figure}[tbp]
    \centering
    \includegraphics[width=\columnwidth]{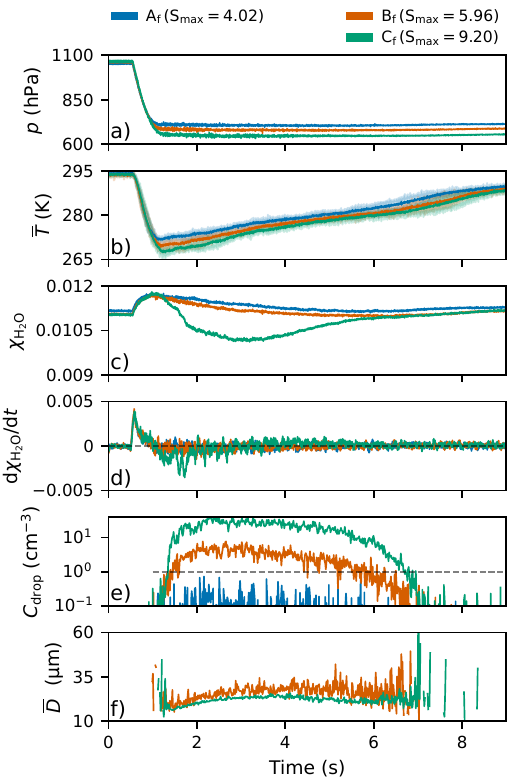}
        \caption{
    Evolution of water vapor and droplet statistics over the course of three expansions performed under forced mixing conditions (fans on) with $S_0=0.50$, corresponding to the color-coded square markers in Figs.~\ref{fig:J_v_Smax} and \ref{fig:J_drops}.
    Experiment A$_\mathrm{f}$ (blue traces) features $\Delta p=\SI{348}{\hecto\pascal}$ and reaches $S_\mathrm{max}=4.02$, below $S_\mathrm{onset}$.   
    Experiment B$_\mathrm{f}$ (orange traces) features $\Delta p=\SI{383}{\hecto\pascal}$ and reaches $S_\mathrm{max}=5.96$, near  $S_\mathrm{onset}$.
    Experiment C$_\mathrm{f}$ (green traces) features $\Delta p=\SI{413}{\hecto\pascal}$ and reaches $S_\mathrm{max}=9.20$, above $S_\mathrm{onset}$.    
    Panels (a)-(f) plot the same quantities reported in Fig.~\ref{fig:ABC}.
    %
    %
    %
    %
    %
    %
    %
    }
    \label{fig:ABC_fans}
\end{figure}

Following the initial evaporation of water from the chamber walls, 
experiment C$_\mathrm{u}$ reaches the highest $S_\mathrm{max}$ and therefore has the highest rate of homogeneous nucleation, per Fig.~\ref{fig:J_v_Smax}b.
Accordingly, we see a significant decrease in $\chi_{\ce{H2O}}$ in this experiment in Fig.~\ref{fig:ABC}c as water vapor condenses into droplets.
We also see $\mathrm{d}\chi_{\ce{H2O}}/\mathrm{d}t$ plunge below zero in Fig.~\ref{fig:ABC}d at  $t=\SI{1}{\second}$, around the same time the cloud chamber reaches its final pressure.
The holography measurements show a steady growth in $C_\mathrm{drop}$ beginning $\sim\SI{0.5}{\second}$ later (Fig.~\ref{fig:ABC}e).
This delay in droplet detection following the expansion is likely due to the fact that droplets must reach diameters above \SI{8}{\micro\metre} for detection with holography.
The results of experiment B$_\mathrm{u}$ evidence similar trends, albeit with much lower $C_\mathrm{drop}$ detected close to the onset of homogeneous nucleation.

We can continue to monitor the evolution of both the droplet size distribution and the water vapor concentration at longer times post-expansion.
Again, trends are easiest to see in the C$_\mathrm{u}$ expansion. 
$C_\mathrm{drop}$ continues to increase for several seconds post-expansion while $\chi_{\ce{H2O}}$ decreases.
$C_\mathrm{drop}$ levels off at \SI{2}{\second}  and $\chi_{\ce{H2O}}$ reaches its minimum at around \SI{6}{\second}.
The system then begins to thermalize with the room temperature chamber walls, leading to evaporation of droplets, a dropping $C_\mathrm{drop}$, and a rising $\chi_{\ce{H2O}}$.

$\overline{D}$ increases post-expansion, and then levels off around \SI{4}{\second}. 
This general trend in $\overline{D}$ is consistent with diffusive droplet growth, as expected for the droplet size regime we are detecting.
Interestingly, $\overline{D}$ continues to hold steady for times past \SI{6}{\second}, even as droplets are evaporating and $C_\mathrm{drop}$ falls.
We acknowledge two explanations for this phenomenon, which likely both contribute in part.
First, the droplet size distribution broadens as the chamber thermalizes.
As the ensemble of droplets evaporates, the smallest droplets shrink below the holography detection threshold and no longer contribute to the $C_\mathrm{drop}$ count.
Meanwhile, the class of larger droplets persists and can therefore bias the detected size distribution towards larger diameters. 
In addition, the inhomogenous mixing paradigm\cite{latham1977laboratory,baker1980influence} -- a ubiquitous phenomenon in cloudy air masses\cite{beals2015holographic,hoffmann2019entrainment,yeom2023cloud} -- is at play here. 
In inhomogeneous mixing, thermodynamic cloud properties like temperature homogenize slowly compared to the time it takes droplets to equilibrate with their local environment, leading to air pockets with distinct droplet concentration and size distributions.
Heat transferred from the chamber walls may therefore preferentially evaporate droplets only in nearby pockets of air, 
leading to little change in $\overline{D}$ at longer waiting times but a significant change in the number of droplets passing through the holography detection volume.
We have also previously invoked inhomogeneous mixing in the REACh chamber to explain a similar relationship observed between $C_\mathrm{drop}$ and $\overline{D}$ in our earlier work on heterogeneous droplet nucleation and growth using phase Doppler anemometry (PDA).\cite{erinin2025droplet} 
PDA features an order-of-magnitude smaller measurement volume than the holography measurement used in the present work, suggesting that inhomogeneous mixing trend persists across spatial scales.

The time-resolved data in Fig.~\ref{fig:ABC}c provides some additional evidence in support of inhomogeneous mixing.
The $\chi_{\ce{H2O}}$ traces for experiment C$_\mathrm{u}$ (green trace) show clear fluctuations in water vapor concentration for the first several seconds post-expansion, pointing to inhomogeneously mixed pockets of air with higher and lower water vapor content passing through the TDLAS measurement volume.
As TDLAS is path-integrated across the chamber diameter, the system is almost inherently insensitive to small-scale variations in water vapor.
The temporal fluctuations observed here are therefore likely driven by macroscopic thermal plumes characteristic of the unforced mixing.
A simple visual inspection of the chamber during an expansion similarly shows regions of brighter and dimmer light scattering along the visible holography beam.

\subsection{The impact of mixing on droplet nucleation and growth}

We finally consider how forced turbulent mixing in the cloud chamber impacts homogeneous nucleation and growth of water droplets.
Characterization of the chamber flow conditions under forced mixing is detailed in our earlier work.~\cite{erinin2025droplet}
While both forced and unforced mixing yield turbulent flow environments, the velocity fluctuations in the chamber are roughly $10\times$ larger under forced mixing conditions than in the unforced case, reaching $\sim \SI{0.8}{m/s}$. 
Forced mixing is therefore expected to enhance small scale variations in the temperature and water vapor fields and accelerate post-expansion thermalization with the room temperature walls.
It should however reduce large scale fluctuations in temperature and water concentration as the overall flow is faster.
We also expect forced mixing to enhance the inhomogeneous mixing processes discussed above in Section \ref{sec:growth} and thereby broaden the droplet size distribution.

Figure \ref{fig:ABC_fans} plots the temporal evolution of droplet formation and growth under forced mixing conditions (fans on) for the same expansions labeled A$_\mathrm{f}$, B$_\mathrm{f}$, and C$_\mathrm{f}$ in Figs.~\ref{fig:J_v_Smax} and \ref{fig:J_drops}.
From visual comparison of Figs.~\ref{fig:ABC} and \ref{fig:ABC_fans}, the same broad trends in $\chi_{\ce{H2O}}$, $C_\mathrm{drop}$, and $\overline{D}$ are evident under both forced and unforced mixing conditions.
However, the timescales of nucleation, condensation, and evaporation are all accelerated in the forced mixing case.
For instance, under unforced mixing, $C_\mathrm{drop}$ holds steady until $t=\SI{7}{\second}$ (Fig.~\ref{fig:ABC}e), whereas the droplet ensemble is nearly fully evaporated by this time under forced mixing (Fig.~\ref{fig:ABC_fans}e).

\begin{figure}[tbp]
    \centering
    \includegraphics[width=\columnwidth]{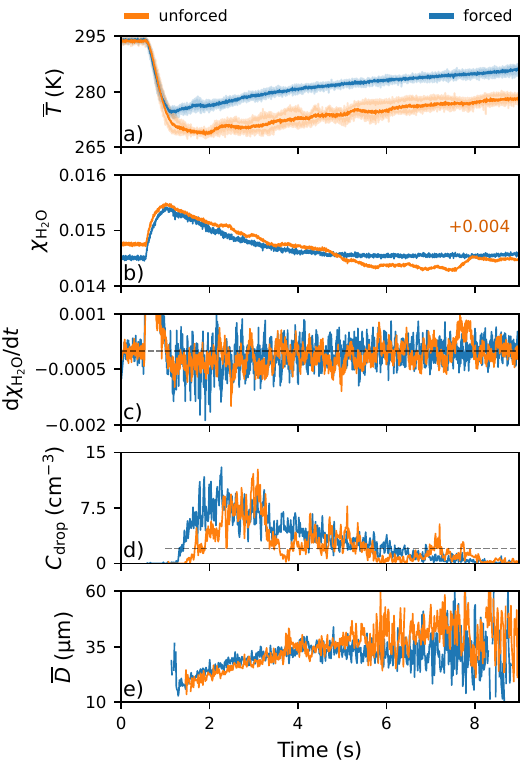}
    \caption{
    Evolution of water vapor and droplet statistics over the course of two expansions featuring unforced mixing (orange traces) and forced mixing (blue traces) that reach similar $S_\mathrm{max}$. 
    The unforced mixing experiment features $S_0=0.50$ and $\Delta p=\SI{378}{\hecto\pascal}$, reaching $S_\mathrm{max}=5.40$ while the forced mixing experiment features $S_0=0.65$ and $\Delta p=\SI{347}{\hecto\pascal}$ and reaches $S_\mathrm{max}=5.31$.
    \textbf{(a)} Unadjusted average temperature reading across the thermocouple array; the shaded envelopes show the coldest and warmest readings.
    \textbf{(b)} Mole fraction of water vapor measured with TDLAS.
    The unforced mixing trace (orange) is shifted by +0.004 so that the maxima visually coincide to account for different $S_\mathrm{0}$.
    \textbf{(c)} $\mathrm{d}\chi_{\ce{H2O}}/\mathrm{d}t$ calculated numerically after smoothing the traces in panel (a). 
    \textbf{(d)} Droplet concentration tracked with holography. 
    The dashed black line indicates the droplet concentration threshold of $\SI{2}{cm^{-3}}$ that distinguishes homogeneously nucleated droplets from activation of background particles.
    \textbf{(e)} Ensemble average droplet diameter tracked with holography. 
    }
    \label{fig:mix_compare}
\end{figure}

To better understand the impact of mixing, Fig.~\ref{fig:mix_compare} plots time traces of $\chi_{\ce{H2O}}$, $C_\mathrm{drop}$, and $\overline{D}$ from a pair of unforced and forced mixing experiments that reach similar $S_\mathrm{max}$ and $C_\mathrm{drop, max}$. 
Full experimental details of these runs are provided in Table S1 of the SM in the rows labeled \textit{u} and \textit{f}.
Figure~\ref{fig:mix_compare}a compares the unadjusted temperature recorded throughout the two expansions.
The unforced and forced mixing expansions reach minimum $T_\mathrm{min,adj}$ temperatures of \SI{255}{\kelvin} and \SI{260}{\kelvin}, respectively.
Under forced mixing, the system spends significantly less time near its minimum temperature and warms evenly with only minor fluctuations in temperature as compared to the unforced case. 
The forced mixing experiment also shows a narrower range in measured temperatures across the thermocouples (compare the widths of the shaded areas in Fig.~\ref{fig:mix_compare}a), suggesting that the fans homogenize temperature gradients inside the chamber on the $\sim\SI{5}{\centi\metre}$ spatial scale of the thermocouple array.
We confirm this by plotting the spatially-resolved temperature time traces across the thermocouple array for these experiments in Fig.~S2 of the SM.
Forced mixing also appears to smooth the evolution of $\chi_{\ce{H2O}}$ in Fig.~\ref{fig:mix_compare}b.



Figures \ref{fig:mix_compare}d,e compare the $C_\mathrm{drop}$ and $\overline{D}$ droplet statistics between forced and unforced mixing. 
At first glance, these statistics are similar: $C_\mathrm{drop}$ reaches maxima of \SI{13}{\per\cubic\centi\metre} in the forced case and \SI{14}{\per\cubic\centi\metre} in the unforced case; at \SI{3}{\second}, before $C_\mathrm{drop}$ decreases significantly in both experiments,
$\overline{D}$ is \SI{29}{\micro\metre} and \SI{25}{\micro\metre} for the forced and unforced cases, respectively.
The growth timescales in $\overline{D}$ are very similar between the two cases, though the time evolution in $C_\mathrm{drop}$ differs. 
In particular, $C_\mathrm{drop}$ begins to increase and peaks markedly earlier under forced mixing.
This trend can be attributed to the existence of more small pockets of air with higher $S$ under forced mixing, leading to rapid $C_\mathrm{drop}$ growth.
$C_\mathrm{drop}$ decreases very smoothly under forced mixing, similar to the trends noted above in temperature and water vapor content. 
The unforced mixing experiment demonstrates large swings in $C_\mathrm{drop}$ as the particles evaporate (see $t=\SIrange{6}{8}{\second}$ in particular), clear evidence of large-scale inhomogeneous mixing. 

\begin{figure}
    \centering
    \includegraphics[width=\columnwidth]{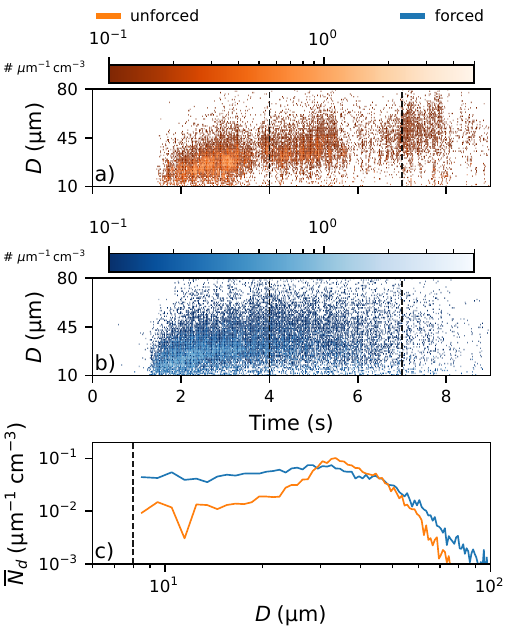}
    \caption{
     Evolution of droplet size distributions for the same two expansions from Fig.~\ref{fig:mix_compare} with unforced (orange traces) and forced mixing (blue traces). 
    \textbf{(a,b)} Droplet size distributions in the unforced and forced mixing cases, illustrating the number of droplets detected in each \SI{500}{\nano\metre} size bin and \SI{1.5}{\milli\second} time bin, normalized by the droplet-size-dependent holographic detection volume and therefore reported in droplets \si{\per\micro\metre} \si{\per\cubic\centi\metre} (see color bar).
    The vertical dashed lines indicate the \SIrange{4}{7}{\second} time window used to report the time-averaged droplet size distribution shown in panel (c).
    \textbf{(c)} The normalized average droplet size distribution recorded in the unforced and forced mixing experiments  averaged over the \SIrange{4}{7}{\second} time window shown in dashed lines in panels (a,b). The vertical dashed line at \SI{8}{\micro\metre} shows the lower reliable size limit of the holographic reconstruction.
    }
    \label{fig:drop_percent}
\end{figure}

Considering the droplet size distribution, rather than just the mean droplet diameter $\overline{D}$, can further inform this discussion.
Figures \ref{fig:drop_percent}a,b plot the evolution in time of the droplet size distribution for the same unforced and forced mixing experiments from Fig.~\ref{fig:mix_compare}.
Figure~\ref{fig:drop_percent}c compares the time averaged droplet size distribution, $\bar{N}_d$ for the two experiments calculated over the \SIrange{4}{7}{\second} time window.
It is clear that forced mixing yields a broader droplet size distribution than the unforced mixing case.
We again ascribe these trends to inhomogeneous mixing: the high degree of turbulence under forced mixing yields small-scale pockets of air with elevated water vapor content or lower temperatures that support growth of the largest droplets -- as well as warmer, dryer air pockets that yield only the tiniest droplets.
The cloud physics literature has invoked similar explanations; see, e.g., prior discussions on broadened size distributions in droplet growth under turbulence\cite{koch2026effect,beals2015holographic,lehmann2009homogeneous} and studies on the evaporation of dense sprays of droplets in a fluctuating vapor field.\cite{villermaux2017fine,pushenko2025diffuselet}

It is clear from these experiments that the spatial scale of the inhomogeneous mixing in the chamber is highly sensitive to flow conditions. 
Unforced mixing leads to large-scale fluctuations that modulate $T$ and $\chi_{\ce{H2O}}$ in time and space, causing large swings in $C_\mathrm{drop}$ over time in the holography data.
Under forced mixing, we see clear evidence for inhomogeneous broadening in the droplet size distributions measured with holography, but no large swings in $C_\mathrm{drop}$ in time.
We conclude that forced mixing must yield inhomogeneously mixed air pockets that are smaller than the holography measurement volume, and certainly smaller than the centimeter-scale spacing of the thermocouples, which read out uniform temperatures under forced mixing. 
It might be of interest in future work to better resolve spatial trends by correlating droplet statistics with their location within the holography mode volume.


\section{Conclusion}
We report on the use of the REACh facility to study the homogeneous nucleation of water vapor.
Using a combination of TDLAS and in-line holography we track droplet nucleation and growth in real time, while simultaneously monitoring the water vapor concentration inside the chamber and extracting the peak saturation ratio reached during each expansion.
Our measurements of droplet nucleation rates show reasonable agreement with CNT.
Both the TDLAS and holography data show clear evidence of inhomogeneously mixed pockets of air in the chamber with variation in water vapor concentration and temperature, which have impacts on homogeneous nucleation and the resulting droplet growth.
We examine different degrees of mixing in the chamber to better understand how droplet nucleation and growth are impacted by these inhomogeneities.
Forced mixing in the chamber appears to accelerate the initial formation of droplets and ultimately broadens the droplet size distribution.
While our laboratory conditions operate at much higher $S$ than occurs in Earth's atmosphere, our results highlight more generally that the large-scale cloud dynamics and turbulence that have major impacts on droplet microphysics are also present during expansions in the REACh facility
Future work is necessary to further investigate and verify these findings across a wider range of turbulent conditions.
In ongoing work, we are broadening our experimental capabilities, including developing a rapid TDLAS readout for temperature, and pursuing further developments in holography processing to track the position of the droplets in three dimensions during the expansion.
We plan to apply these methodologies to study heterogeneous nucleation of liquid droplets and heterogeneous ice nucleation on seeding aerosol particles.
\section*{Supplementary Material} \vspace{-2ex}
\noindent
See the Supplementary Material for the details of temperature correction (Section SI), spatially resolved temperature evolution during expansions (Fig.~S2), and a table (Table~S1) including key statistics from every experiment studied in this report. 
Also included are two processed holography videos showing interference caused by too high droplet counts, shown as Fig.~S3 and Fig.~S4. 

\begin{acknowledgments}
The research described in this paper was conducted under the Laboratory Directed Research and Development (LDRD) Program at Princeton Plasma Physics Laboratory, a national laboratory operated by Princeton University for the U.S. Department of Energy under Prime Contract No. DE-AC02-09CH11466. The United States Government retains a non-exclusive, paid-up, irrevocable, world-wide license to publish or reproduce the published form of this manuscript, or allow others to do so, for United States Government purposes.
This work was supported by the Simons Foundation through Award No.\ SFI-MPS-SRM-00011998, and by internal Princeton University funds from the Princeton Catalysis Initiative and the High Meadows Environmental Institute’s Climate and Energy Grand Challenge Award granted to L.D. and M.L.W. M.L.W. also acknowledges support from the Packard Foundation. We acknowledge helpful discussions with Raymond Shaw.
\end{acknowledgments}

\section*{Author Declarations}

\subsection*{Conflict of Interest} \vspace{-2ex}
\noindent The authors have no conflicts to disclose.

\subsection*{Author Contributions} \vspace{-2ex}
\noindent 
\textbf{Cole R. Sagan}: Conceptualization (equal); Data curation (lead); Formal analysis (lead); Investigation (equal); Methodology (equal); Validation (equal); Visualization (equal); Writing – original draft (lead); Writing – review \& editing (equal). 
\textbf{Gwenore Pokrifka}: Conceptualization (equal); Data curation (supporting); Formal analysis (supporting); Investigation (equal); Methodology (equal); Validation (equal); Visualization (equal); Writing – review \& editing (equal).
\textbf{Samuel Koblensky}: Data curation (supporting); Formal analysis (supporting); Methodology (supporting); Validation (supporting); Visualization (supporting); Writing – review \& editing (equal).
\textbf{Martin A. Erinin}: Data curation (supporting); Formal analysis (supporting); Investigation (equal); Writing – review \& editing (supporting). 
\textbf{Ilian Ahmed}: Data curation (supporting); Formal analysis (supporting); Investigation (equal); Writing – review \& editing (supporting). 
\textbf{Nadir Jeevanjee}: Formal analysis (supporting); Investigation (supporting); Writing – review \& editing (supporting). 
\textbf{Luc Deike}: Conceptualization (equal); Funding acquisition (equal); Investigation (equal); Methodology (equal); Project administration (equal); Resources (equal); Supervision (equal); Writing – review \& editing (equal).
\textbf{Marissa L. Weichman}: Conceptualization (equal); Funding acquisition (equal); Methodology (equal); Investigation (equal); Project administration (equal); Resources (equal); Supervision (equal); Writing – review \& editing (equal).

\section*{Data Availability} \vspace{-2ex}
\noindent The data that support the findings of this study are available from the corresponding authors upon request.

\section*{References} \vspace{-2ex}
\bibliography{main_bib}

\end{document}